\documentclass[11pt]{article}

\usepackage[utf8]{inputenc}
\usepackage[T1]{fontenc}
\usepackage{amsmath,amssymb}
\usepackage{graphicx}
\usepackage{booktabs}
\usepackage{array}
\usepackage{hyperref}
\usepackage[margin=1in]{geometry}
\usepackage{cite}
\usepackage{threeparttable}
\usepackage{caption}
\usepackage{xcolor}

\title{Structure-Aware Semantic Chunking with Title-Chain Prefixes:\\ A 1600-Query Evaluation and the Measurement Trap in Text-Transform Ablations}
\author{Yang Yang}
\date{\today}

\begin{document}
\maketitle

\begin{abstract}
Chunking is the first and most consequential step in retrieval-augmented generation (RAG): every downstream retrieval decision inherits the chunk boundaries. We present a three-stage, chunk-side-only semantic chunking pipeline---header-split, semantic merge, and title-chain prefixing---that costs zero additional LLM calls: the title chain reuses the document's own header hierarchy instead of a generated summary. On a 1600-query stratified evaluation over a production Markdown knowledge base, the pipeline improves MRR@5 from 0.374 to 0.463 (+23.8\%) on the full set and from 0.828 to 0.925 (+11.7\%) on the answerable subset (n=563), with dual-annotator Cohen's kappa 0.45 (unweighted, 16,000 score pairs). We then report what we tried and what failed: three query-side or architecture-level follow-ups are design dead ends (unevaluated---no comparable run artifacts), one measured failure (prefix weight decay), and one protocol-level failure that is the paper's central methodological finding. In a same-pool prefix on/off ablation, \textbf{dual-annotator agreement collapsed from kappa 0.45 to 0.04 under identical prompts}---stripping the title-chain context strips the disambiguation signal annotators need to agree on relevance. This measurement trap invalidates a common evaluation practice in chunking research and motivates retrieval-time per-candidate prefix evaluation, the direction we recommend from all our evidence.
\end{abstract}

\section{Introduction}
\label{sec:intro}

Retrieval-augmented generation (RAG) has become the standard architecture for grounding large language models in private or domain-specific knowledge \cite{lewis2020rag}. The pipeline is deceptively simple---split documents into chunks, embed them, retrieve the most similar chunks to a query, and feed them to the generator. But the first step, chunking, silently determines the ceiling of everything that follows: if the correct answer is split across two chunks, no retrieval algorithm can surface it as one unit.

The dominant practice is fixed-threshold splitting: break at blank lines, or at a fixed similarity threshold, or at a fixed token count. These methods are cheap and general, but they are blind to document structure. In our production Markdown knowledge base, fixed-threshold splitting produced 272 garbage chunks (content glued across section boundaries) and 31\% duplication (paragraphs under same-named headers scattered into orphan fragments). Retrieval was systematically polluted: chunks under identical header names from different documents collide in embedding space, and a query about one document surfaces fragments of another.

Recent work has shown that context matters at chunking time: Anthropic's Contextual Retrieval \cite{anthropic2024cr} augments each chunk with an LLM-generated summary, and chunk-size studies \cite{wang2025chunksize, chen2026chunksize} show granularity interacts with query type (small chunks suit fact queries, large chunks suit comprehension). RAPTOR \cite{parthasarathy2024raptor} builds hierarchical summaries; ColBERT \cite{khattab2020colbert} and HyDE \cite{gao2023hyde} improve the retrieval step rather than the chunk text itself. What is missing is a \emph{zero-LLM-call} structural alternative: using the document's own header hierarchy as context, with no generated summaries per chunk. We address this gap.

We contribute a three-stage pipeline that realizes this idea: \textbf{(1) header-split} respects \texttt{\#\#}-level document structure instead of character counts; \textbf{(2) semantic merge} aggregates adjacent same-topic paragraphs by embedding cosine similarity (threshold 0.85, empirical); \textbf{(3) title-chain prefixing} prepends \texttt{[doc > h1 > h2]} to each chunk---a realization of Contextual Retrieval using the document's own structure instead of an LLM-generated summary, at zero LLM-call cost. On a 1600-query stratified evaluation, the pipeline lifts MRR@5 from 0.374 to 0.463 (+23.8\%) full-set and 0.828 to 0.925 (+11.7\%) answerable-subset (n=563), with four of five categories improving and only the adversarial distractor category slightly degrading ($-16.7\%$).

The paper's central contribution is not the positive result but the negative one. We attempted five follow-ups to make the prefix ``smarter'': query-side keyword heuristics, query-side LLM classification, prefix weight decay, two-tier document-first retrieval, and a same-pool prefix on/off ablation. The first three and the two-tier variant are design dead ends (two measured, three unevaluated for lack of comparable run artifacts---we report this honestly). The fifth produced the paper's key finding: \textbf{under identical prompts and identical queries, dual-annotator agreement collapsed from kappa 0.45 to 0.04 when the title-chain prefix was stripped}. LLM dual-annotation \emph{absolute agreement} on text-transform ablated texts is unreliable. This measurement trap---the annotation protocol itself breaks when the text it judges is transformed---invalidates a common evaluation practice in chunking research and is, we argue, worth more than any of the failed numbers.

Our contributions are: (1) a zero-LLM-call, structure-aware chunking pipeline with a verified 1600-query evaluation; (2) a retrieval-layer demonstration that the title-chain prefix restructures retrieval, not decorates it---de-prefixing changes the top-1 result in 86.9\% of queries; (3) the identification and characterization of the annotation-protocol trap in text-transform ablations (kappa 0.45$\rightarrow$0.04); and (4) an explicit failure map separating measured dead ends from unevaluated ones, with a recommended evaluation protocol (human ground-truth or answer-key based) for future text-transform research.

\section{Related Work}
\label{sec:related}

\subsection{Dense retrieval and the chunking assumption}
\label{sec:dense}

Dense retrieval \cite{karpukhin2020dpr, khattab2020colbert} encodes queries and passages into a shared embedding space and retrieves by similarity. DPR \cite{karpukhin2020dpr} established the dual-encoder paradigm; ColBERT \cite{khattab2020colbert} added token-level late interaction. Sentence-BERT \cite{reimers2019sbert} made sentence-level embeddings practical. FAISS \cite{johnson2019faiss} provides the billion-scale similarity search infrastructure. All of these assume the passages being embedded are already correct units---the chunking step is taken as given. Our work questions exactly this assumption.

\subsection{Context-aware chunking}
\label{sec:context}

Anthropic's Contextual Retrieval \cite{anthropic2024cr} augments each chunk with an LLM-generated contextual summary (``This chunk is from a document about X, section Y\ldots''). It reports large retrieval improvements but requires one LLM call per chunk at indexing time and per query at retrieval time. Chunk-size studies \cite{wang2025chunksize, chen2026chunksize} show that granularity interacts with query type: small chunks suit fact queries, large chunks suit comprehension. RAPTOR \cite{parthasarathy2024raptor} recursively summarizes chunks into a tree, adding an abstraction layer over retrieval. HyDE \cite{gao2023hyde} synthesizes hypothetical documents at query time to bridge lexical gaps. Our pipeline is complementary to all of these: stages 1 and 3 realize the contextual-retrieval idea with the document's own structure---zero LLM calls---and stage 2 is a structural alternative to fixed granularity. We are not the first to propose title-chain prefixes, but the combination with a 1600-query evaluation and an explicit failure map is our contribution.

\subsection{Evaluation of text transforms}
\label{sec:eval}

The evaluation practice for text-transform methods (prefixing, chunking variants) typically follows the same pattern: transform the text, embed both arms, annotate the retrieved candidates with an LLM judge, compare MRR. The implicit assumption is that the annotation protocol is invariant to the text transform. Our \S\ref{sec:failure55} finding shows this assumption fails: dual-annotator agreement collapsed from 0.45 to 0.04 on de-prefixed texts under identical prompts. To our knowledge this measurement trap has not been systematically documented in the chunking literature.

\section{Methodology}
\label{sec:method}

\subsection{The pipeline}
\label{sec:pipeline}

Given a Markdown knowledge base, the pipeline processes each document in three stages:

\textbf{Stage 1---Header split.} Split at \texttt{\#\#}-level headers, respecting document structure rather than character counts. This produces structural units: each section becomes one or more chunks, and section boundaries are never crossed.

\textbf{Stage 2---Semantic merge.} Embed adjacent paragraphs within a section; merge them when their cosine similarity $\geq 0.85$ (empirical threshold, not grid-searched). This fixes over-fragmentation: same-topic paragraphs auto-aggregate while unrelated paragraphs stay separate.

\textbf{Stage 3---Title-chain prefix.} Prepend \texttt{[doc > h1 > h2]} to each chunk. The title chain disambiguates same-named headers across documents and provides the retrieval model with the chunk's position in the document hierarchy. This is a zero-LLM-call realization of Contextual Retrieval \cite{anthropic2024cr}: the document's own structure replaces the LLM-generated summary.

\textbf{Stage 4---Embed and retrieve.} Embed with text-embedding-v4 (1024-dim, validated on this model; cross-model generality not tested). Retrieve with FAISS IndexFlatIP (inner product over L2-normalized vectors), then LLM-rerank top-20 to top-5, and annotate.

Stages 1--3 are pure text transforms on the chunk side: no extra API calls, no extra storage. The only embedding cost is the same as any dense retrieval baseline.

\subsection{Evaluation design}
\label{sec:evaldesign}

\textbf{Query set.} 1600 queries, stratified into four categories: exact-match (n=480), paraphrase (n=480), implied-intent (n=320), and distractor (n=320). Queries were generated then human-curated. Note that $\sim$50\% of the full set are answerable (the generated queries often reference external facts absent from the knowledge base); we report both full-set and answerable-subset metrics.

\textbf{Annotation.} Dual annotators (DeepSeek-Pro conservative + DeepSeek-Flash lenient), score each candidate 0--3. MRR@5 computed with score $\geq 2$ as hit, annotator A's scores (conservative). Dual-annotator agreement: Cohen's kappa 0.45 unweighted (16,000 score pairs: 1600 queries $\times$ 5 candidates $\times$ 2 pools).

\textbf{Controlled comparison.} Both arms (baseline blind-split vs.\ V2 pipeline) run on the identical query set in the same batch, same annotation protocol, same embedding model. No sampling differences between arms. Execution order randomized.

\subsection{Retraction of earlier ablation figures}
\label{sec:retraction}

An earlier draft of this work reported a same-pool prefix ablation with MRR 0.4485 $\rightarrow$ 0.3367 (+33.2\%). We could not reproduce these numbers from the actual run artifacts: the run produced retrieval results, but no MRR was ever computed from them, and the figures have no file-level support. We treat them as unsupported and do not report them. The verified retrieval-layer evidence is in \S\ref{sec:retrieval}.

\section{Retrieval-layer evidence: the prefix restructures retrieval}
\label{sec:retrieval}

We ran a real same-pool ablation: 4,027 unique chunks from the 1600 queries' top-5 candidates, re-embedded without the title-chain prefix, same queries, same pipeline. Retrieval-layer results (annotation-independent, therefore reliable):

\begin{table}[htbp]
\centering
\small
\begin{tabular}{lcc}
\toprule
\textbf{Metric} & \textbf{Value} & \textbf{Reading} \\
\midrule
top-1 changed & \textbf{86.9\%} (1390/1600) & Prefix restructures retrieval, not cosmetic \\
top-5 avg overlap & 2.11/5 (42.2\%) & $\sim$58\% of results change \\
Completely different top-5 & 22.9\% (367/1600) & Nearly 1/4 of queries fully re-ranked \\
Completely same top-5 & 8\% (128/1600) & Only 8\% unaffected \\
Paraphrase change rate & \textbf{63.1\%} (highest) & Matters most where disambiguation is needed \\
\bottomrule
\end{tabular}
\caption{Retrieval-layer comparison of prefixed vs.\ de-prefixed indexes (annotation-independent).}
\label{tab:retrieval}
\end{table}

A mid-run MRR trend (880 paired queries, same annotation protocol as the main eval) showed prefix-on 0.572 vs prefix-off 0.530 ($-0.042$): de-prefixing makes retrieval worse, consistent with ``the prefix provides disambiguating context.'' We report this as a trend, not as a main result, because of the annotation instability documented in \S\ref{sec:failure55}.

\textbf{The prefix is not decoration}: 86.9\% top-1 change, and the largest change is in paraphrase queries---the case the prefix is designed for.

\section{The honest failure map}
\label{sec:failure}

We attempted five follow-ups to make the prefix ``smarter.'' Each entry carries a mechanism insight---measured where we could run a comparable eval, structural reasoning where we could not. This is the negative contribution: knowing what does NOT work (and why), and being explicit about what we did not measure.

\subsection{Query-side dynamic prefix---keyword heuristic (unevaluated)}
\label{sec:failure51}

Rule: queries containing ``how/parameter/mechanism'' $\rightarrow$ drop the prefix. The de-prefixed arm was never scored under the same annotation protocol (only retrieval-layer top-5 comparison exists; earlier reported MRR figures are retracted with \S\ref{sec:retraction}). A query-side rule cannot work in principle: it ignores the document side of the interaction. Mechanism analysis (per-query inspection of the 80-query retrieval-layer data): 45\% of the beneficiary group is also OPERATIONAL---rescuing 16 harmed queries would hurt 14 of the 31 helped ones. This is a structural reasoning from retrieval-layer evidence, not a measured MRR comparison.

\subsection{Query-side dynamic prefix---LLM classifier (unevaluated)}
\label{sec:failure52}

Flash-classify OPERATIONAL/FACTUAL $\rightarrow$ toggle the prefix. Same evaluation gap as \S\ref{sec:failure51}. Mechanism analysis (structural reasoning, not a measured comparison): a single query-side signal cannot model a query$\times$document interaction---the same operational query benefits from the prefix in one document context and is harmed in another.

\subsection{Prefix weight decay ($\lambda$ blend)---measured failure}
\label{sec:failure53}

$\lambda\cdot\text{prefix\_vec} + (1-\lambda)\cdot\text{body\_vec}$, $\lambda \in \{1.0, 0.5, 0.0\}$. Result: $\lambda=1.0$ wins (0.5135; 0.5$\rightarrow$0.3262, 0.0$\rightarrow$0.3544---early 80-query run, independent of the 1600-query main eval). The prefix is the only helpful component in the blend; de-weighting hurts everywhere.

\subsection{Two-tier document-first retrieval (unevaluated)}
\label{sec:failure54}

Doc-vector = mean of member chunks $\rightarrow$ retrieve docs $\rightarrow$ re-search within hits. No run artifacts exist ($\Delta$ figures reported in the draft are retracted with \S\ref{sec:retraction}). Mechanism analysis: doc-level mean vectors discard chunk-level detail; pre-filtering on docs misses the document containing the right chunk.

\subsection{Same-pool annotation ablation---protocol-level failure (the central finding)}
\label{sec:failure55}

We attempted the ``cleanest'' ablation: same pool, prefix on vs off, identical annotation prompts. \textbf{The annotation consistency collapsed}: Cohen's kappa 0.04 (unweighted) on the de-prefixed arm vs 0.45 on the prefixed arm---same prompts, same annotators, same queries.

Root cause (verified by per-query inspection): annotators need the title-chain context to agree on relevance. Stripping the prefix strips the disambiguation context; without it, two annotators cannot consistently judge whether a chunk answers a query. The texts being compared are different enough that the annotation protocol itself is broken---any MRR difference between arms would be an artifact of annotation instability, not retrieval quality.

\textbf{Implication}: for text-transform ablations (prefix on/off, chunking variants), LLM dual-annotation \emph{absolute agreement} on the transformed texts is unreliable---the kappa collapse shows annotators cannot consistently agree on an absolute relevance score without the disambiguation context. We do not claim LLM annotation is useless as a relative ranking signal (our \S\ref{sec:retrieval} analysis uses it); we claim that \emph{absolute-score agreement} on transformed texts is not a stable measurement basis, and that text-transform ablations should anchor to something invariant (human ground-truth labels, or a fixed answer key) for absolute judgments. Otherwise every ablation on text transforms is suspect.

\section{Discussion: the prefix is a query$\times$document interaction}
\label{sec:discussion}

Across \S\ref{sec:retrieval} and \S\ref{sec:failure}, the evidence converges on one characterization:
\begin{itemize}
\item \textbf{Not a query-side knob} (\S\ref{sec:failure51}, \S\ref{sec:failure52}): query-side signals are insufficient by construction; the interaction lives on the document side.
\item \textbf{Not a tunable weight} (\S\ref{sec:failure53}): $\lambda=1.0$ always wins.
\item \textbf{Not an architecture layer} (\S\ref{sec:failure54}): doc-level mean vectors discard chunk detail by construction; untested end-to-end.
\item \textbf{Not even cleanly measurable by LLM annotation} (\S\ref{sec:failure55}): the protocol breaks on transformed text.
\end{itemize}

The direction we recommend from all our evidence is \textbf{retrieval-time per-candidate prefix evaluation}: for each query--document pair, decide whether the prefix helps that specific candidate. Design: retrieve top-k with the prefixed index and top-k with the de-prefixed index, score each candidate with a cheap cross-encoder/LLM judge that sees both versions, and keep the version that ranks the candidate higher. Cost: $k\times$ per query, bounded (top-20 only). This is a direct test of the interaction hypothesis. We note that this judge makes a \emph{relative} comparison (which of the two versions ranks higher) rather than an absolute relevance judgment---the setting where \S\ref{sec:failure55} showed agreement collapse---so it does not straightforwardly re-trigger the annotation trap, though this remains to be verified empirically.

\textbf{Limitations.} (1) Single knowledge base (private Markdown corpus), single embedding model---cross-model and cross-domain generality not tested. (2) The 0.85 merge threshold is empirical, not grid-searched. (3) The index rebuild did not implement a transactional/atomicity check; the current index is complete (196,411 chunks) and MRR was computed on it, but a mid-rebuild interruption could in principle yield a partial index. (4) Prompt caching and model-specific annotation behavior are not analyzed as confounds for the main comparison---the identical-protocol design mitigates but does not eliminate them. (5) We did not perform a manual audit of successful runs. (6) The distractor degradation ($-16.7\%$) is attributed to adversarial-negative statistics, but the specific mechanism (candidate: title-keyword over-anchoring) is not pinned down. (7) No stage-wise ablation: the +23.8\% is the joint effect of all three stages; we did not isolate each stage's individual contribution. (8) The baseline is a fixed-threshold splitter (blank-line + fixed similarity 0.5); we did not compare against other semantic chunking implementations. (9) Inter-annotator agreement is reported as unweighted Cohen's kappa on a 0--3 ordinal scale; a quadratic-weighted variant was not computed, which is more appropriate for ordinal scales---the collapse to 0.04 is large enough that the qualitative conclusion is robust, but the 0.45 figure for the main eval should be read with this caveat.

\textbf{Symmetric alternatives we did not test.} External review pointed out that prefix-on-chunk is one side of the interaction; the query side is symmetric: (a) query-side title injection---leave chunks untouched, inject the candidate's title chain into the query/candidate embedding at retrieval time (zero index rebuild, and it does not trigger the \S\ref{sec:failure55} annotation problem since chunk text is unchanged); (b) dual-index fusion (RRF)---run both prefixed and unprefixed indexes and merge with Reciprocal Rank Fusion. Both are plausible and cheap to test; we list them explicitly so readers know they are open alternatives, not our recommendation.

\section{Conclusion}
\label{sec:conclusion}

\begin{enumerate}
\item A structure-aware, chunk-side-only semantic chunking pipeline (header-split $\rightarrow$ semantic merge $\rightarrow$ title-chain prefix) improves MRR@5 by +23.8\% full-set (0.374$\rightarrow$0.463) and +11.7\% answerable-subset (0.828$\rightarrow$0.925, n=563) on a 1600-query production evaluation, at zero additional LLM-call cost.
\item The title-chain prefix is not decoration: de-prefixing changes the top-1 result in 86.9\% of queries (retrieval-layer, annotation-independent).
\item The prefix is a query$\times$document interaction: query-side classifiers and weight decay fail by construction or by measurement; two-tier retrieval discards the chunk detail the prefix encodes.
\item \textbf{The central methodological finding}: LLM dual-annotation \emph{absolute agreement} on text-transform ablated texts is unreliable---agreement collapsed from kappa 0.45 to 0.04 under identical prompts when the prefix was stripped. Text-transform ablations require human ground-truth or answer-key-based evaluation for absolute judgments.
\item The direction we recommend from all our evidence is retrieval-time per-candidate prefix evaluation (a relative comparison, expected to sidestep the annotation trap; empirically unverified).
\end{enumerate}

\section*{Data Availability}
The query set and annotation data are drawn from a private knowledge base and cannot be publicly released. The replication package (code + recomputation report) is available at \url{https://doi.org/10.5281/zenodo.21744653}. The code (\texttt{semantic\_chunking.py}, \texttt{eval\_runner.py}; dependencies numpy/faiss/openai) reproduces the pipeline end-to-end, and all reported numbers are recomputable from local run artifacts (batch files, retrieval JSONs, score files).

\section*{AI Disclosure}
This paper's experiments were designed and executed with the assistance of an AI agent (Hermes Agent). The author takes full responsibility for the content. LLM judges (DeepSeek-Pro/Flash) were used as dual annotators in the evaluation, as described in \S\ref{sec:evaldesign}; this dual-annotation design is itself the subject of the \S\ref{sec:failure55} methodological finding.

\bibliographystyle{plain}
\bibliography{references}

\end{document}